\documentstyle[11pt,aaspp4]{article}

\begin{document}

\received{}
\revised{}
\accepted{}

\lefthead{Borissova et al.}
\righthead{The outer-halo globular cluster NGC 6229}

\slugcomment{To appear in The Astronomical Journal (Feb. 1997 issue)}

\singlespace

\title{Stellar Photometry of the Globular Cluster NGC 6229.
I. Data Reduction and Morphology of the Brighter Part of the CMD}

\author{J.~Borissova \altaffilmark{1} }
\authoraddr{leonid@bgphys.phys.uni-sofia.bg}
\author{M.~Catelan \altaffilmark{2} }
\authoraddr{catelan@stars.gsfc.nasa.gov}
\author{N.~Spassova \altaffilmark{1} }
\authoraddr{neda@bgearn.acad.bitnet}
\and
\author{A.~V.~Sweigart \altaffilmark{2} }
\authoraddr{sweigart@bach.gsfc.nasa.gov}

\altaffiltext{1}{
   Institute of Astronomy, Bulgarian Academy of Sciences,
   72~Tsarigradsko chauss\`ee, BG\,--\,1784 Sofia, Bulgaria
   }
\altaffiltext{2}{
   NASA/Goddard Space Flight Center, Code 681, Greenbelt,
   MD 20771, USA
   }

\begin{abstract}
$BV$ CCD photometry of the central ($1.5\arcmin \times 2.0\arcmin$) part of the
mildly concentrated outer-halo globular cluster NGC 6229 is presented. The data
reduction in such a crowded field was based on a wavelet transform analysis. 
Our larger dataset extends the previous results by Carney et al. (1991, AJ, 
101, 1699) for the outer and less crowded fields of the cluster, and confirms
that NGC 6229 has a peculiar color-magnitude diagram for its position in the 
Galaxy. In particular, NGC 6229's horizontal branch (HB) presents several 
interesting features, among which stand out: a well populated and very extended
blue tail; a rather blue overall morphology, with 
$(B-R)/(B+V+R) = 0.24\pm 0.02$; a bimodal color distribution, resembling those 
found for NGC 1851 and NGC 2808; and gaps on the blue HB. NGC 6229 is the 
first bimodal-HB cluster to be identified in the Galactic outer halo. A low 
value of the $R$ parameter is confirmed, suggestive of a low helium abundance 
or of the presence of a quite substantial population of extreme HB stars 
fainter than our photometric limit ($\simeq 2.5$ mag below the RR Lyrae level 
in $V$). Twelve new possible variable stars were found in the central part of 
the cluster. The morphology of the red giant branch (RGB) also seems to be 
peculiar. In particular, the RGB luminosity function ``bump" is not a prominent
feature and has only been tentatively identified, on the basis of a comparison 
with a previously reported detection for M3 (NGC 5272). Finally, we compare the
properties
of NGC 6229 with those for other outer-halo globular clusters, and call 
attention to what appears to be a bimodal HB distribution for the outer-halo 
cluster population, where objects with very red {\em or} very blue HB types 
are much more frequently found than clusters with intermediate HB types.

\end{abstract}


\clearpage

\section{Introduction}
The globular cluster (GC) NGC 6229 (C1645+476) is one of the most remote objects
associated with the Galaxy, lying about 28~kpc from the Galactic center
(Harris 1996). The first color-magnitude diagram (CMD) for this low-reddening
[$E(\bv ) \simeq 0.01$ mag; Harris 1996] cluster was presented by Cohen (1986),
who suggested that the horizontal branch (HB) of NGC 6229 is unusually red for
its metallicity (${\rm [Fe/H]} = -1.44$; Harris 1996). Subsequently, Carney et
al. (1991, hereinafter CFT91) obtained extensive $BV$ CCD photometry for the
brighter part of the CMD of NGC 6229. In their survey, CFT91 investigated the
external regions of the cluster beyond $0.8\arcmin$ from the cluster center.
In contrast with the results by Cohen, they detected
a substantial blue HB population, suggesting that the HB of NGC 6229
is instead quite bluer than the average for the outer-halo globulars, where
substantially well-populated extended blue HBs are not usually found.
The available data also
suggest a relative paucity of RR Lyrae variables in comparison with both the
blue and red HB stars in this cluster. A similar effect seems to be
present in the strongly bimodal-HB clusters NGC 2808 (Ferraro et al. 1990)
and NGC 1851 (Walker 1992; Saviane et al. 1996). An explanation for the
origin of the bimodal HB distributions in these clusters is expected to
have a major impact upon our understanding of the second-parameter phenomenon
(e.g., Rood et al. 1993; Stetson et al. 1996). For these reasons,
it is especially important to increase the number of HB stars in the
CMD of NGC 6229.

        This paper presents the first CCD photometry of the central
($1.5\arcmin\times 2.0\arcmin$) region of NGC 6229 performed in
Johnson's ($BV$) photometric system. In Sect. 2, we describe our observations
and explain the adopted non-standard data reduction procedure; an analysis of
the errors is given in Sect. 3; the red giant branch (RGB) is described and
discussed in Sect. 4; the morphology of NGC 6229's HB is
addressed in Sect. 5; the existence of radial population gradients is discussed
in Sect. 6; and, finally, in Sect. 7 the characteristics of NGC 6229 are
compared with those for other outer-halo clusters.
In a subsequent paper (Catelan et al.
1996), we shall present a theoretical analysis of the CMD morphology of
this cluster, with particular emphasis on the reproduction of its HB
morphology by means of synthetic models based upon the latest evolutionary
prescriptions for HB stars and RR Lyrae variables.

\section{Observations and data reduction}

\subsection{Observations}
The observations of NGC 6229 were obtained with the 2m Ritchey--Chr\'etien
telescope of
the Bulgarian National Astronomical Observatory. A set of CCD frames in the
$BV$ system was taken with the SBIG Model ST$-$6 camera. The detector size is
$375\times 242$ pixels, the scale is $0.30\arcsec\times 0.34\arcsec$ per pixel
and the image size is about $1.5\arcmin\times 2.0\arcmin$. This camera was
kindly granted by the EAS/ESO program in support of astronomy in the
Central/Eastern Europe countries. Exposure times ranged from 300 to 900~s in
$V$ and from 900 to 1200~s in $B$.

        During  the  nights of the observations, the seeing was stable with
a measured stellar point-spread function (PSF) on the raw frames of about
$0.8\arcsec$ FWHM. The mean airmass over the duration of each exposure was 
between 1.1 and 1.2.

\subsection{Wavelet transform analysis}

        The preliminary reductions of the CCD frames, including bias subtraction
and flat fielding, were carried out using the standard {\sc IRAF} data reduction
package.

        In the central part of such a concentrated cluster as NGC 6229
($c\simeq 1.61$; Trager et al. 1993), crowding prevented accurate photometry
from being obtained. However, by using the positions of stars determined from
frames after noise suppression using a wavelet transform analysis, we were able
to obtain reasonable photometry at and above the HB level.
Similar techniques had previously been employed by other authors
(e.g., Auri\`ere \& Coupinot 1989) to obtain improved photometric data for
stars in the dense fields that characterize concentrated GCs.

As described by Coupinot et al. (1992), the wavelet transform can be regarded
as a set of ``filters" using a numerical mask with a varying size. It leads to
the decomposition of the image into a set of maps exhibiting structures at
certain scales. In the present application, we used the method to detect and
localize unresolved stars in the core of NGC 6229. We assumed that both the
input images and the PSF can be characterized by gaussian profiles in order
to fit better the seeing-induced core of the stellar image.

Following the method described by  Murtagh et al. (1995), we
carried out a discrete convolution of the image using a filter based on the
$B_3$ spline interpolation. The $5\times 5$ mask was used. The hypothesis of a
white gaussian noise was adopted to determine the noise of the image.
By means of the standard photometric package {\sc DAOPHOT} (Stetson 1987), we
were then able to determine the positions of the stars on the resolved image.

\subsection{Numerical simulations}

In order to check the position accuracy, a set of numerical simulations was
carried out. The first such simulation involved stellar objects in a non-crowded
field. In this experiment, 10 different sets of 500 randomly placed point
sources and a gaussian background were laid out in a grid of $375\times 242$
pixels. The magnitudes of the point sources were random values ranging from
15 to 20 mag. The PSF was a gaussian function with a FWHM equal to 3 pixels.
After wavelet transforming the test images, the {\sc DAOPHOT} fitting
package was used to determine the positions of the
point sources in the ``original" and ``resolved" images. The mean differences
between the positions of the point sources in the ``original" and ``resolved"
images are $\Delta X = 0.002$ ($\sigma = 0.015$) and $\Delta Y = 0.005$
($\sigma = 0.016$) in $X$ and $Y$ pixel coordinates, respectively.

A second simulation was made for stellar objects in a crowded field. We
built test images with the same characteristics as described above, but added a
``concentration" parameter equivalent to King's (1966) concentration parameter
$c$. In this case, we assume $c = 1.6$.
The mean differences  are $\Delta X = 0.01$ ($\sigma = 0.16$) and
$\Delta Y = 0.04$ ($\sigma = 0.16$) in $X$ and $Y$  coordinates, respectively.

\begin{figure}[t]
\plotfiddle{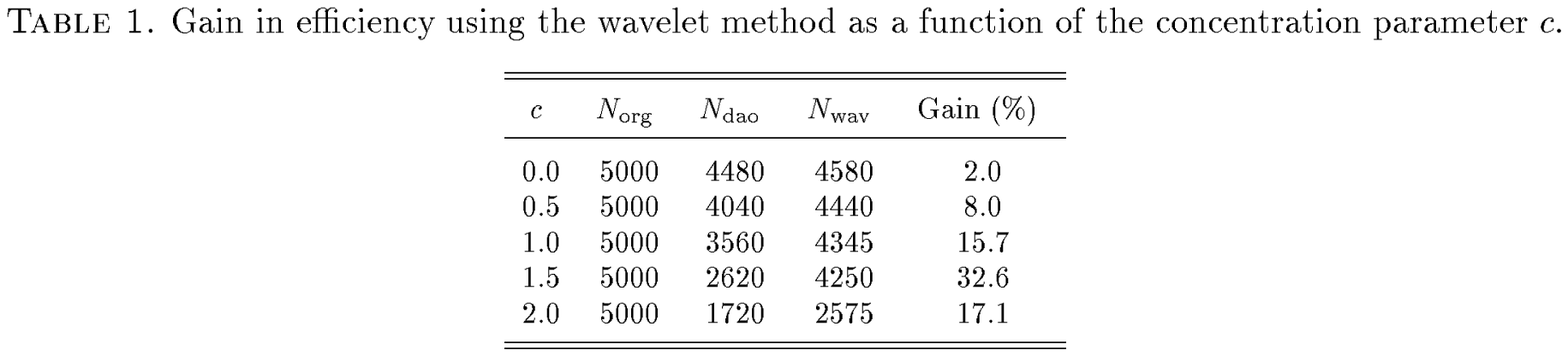}{1.5in}
{0}{90}{90}{-272}{-295}
\end{figure}

A third simulation was made in order to test the efficiency of the wavelet
method as a function of the degree of central concentration. For this purpose,
some test images were built with different concentration parameters $c$, again
assuming a random distribution of magnitudes between 15 and 20 mag. Table 1
summarizes the results. The entries in Table 1 are as follows: $c$ is the
concentration parameter, ranging from 0 (randomly placed point sources) to 2
(highly concentrated cluster); $N_{\rm org}$ is the number of point sources
placed on the ``original" frame; $N_{\rm dao}$ is the number of point sources
determined by {\sc DAOPHOT} on the ``original" frame; and $N_{\rm wav}$
is the number of point sources determined by {\sc DAOPHOT} on the ``resolved"
(wavelet-transformed) image. The last column shows the gain in efficiency,
defined as ${\rm Gain} = (N_{\rm wav}-N_{\rm dao})/N_{\rm org}$, obtained by
using the wavelet transform method.
The method is clearly most useful for medium crowded fields, and our tests
show that this is the case over the entire range of magnitudes.

\begin{figure}[b]
\plotfiddle{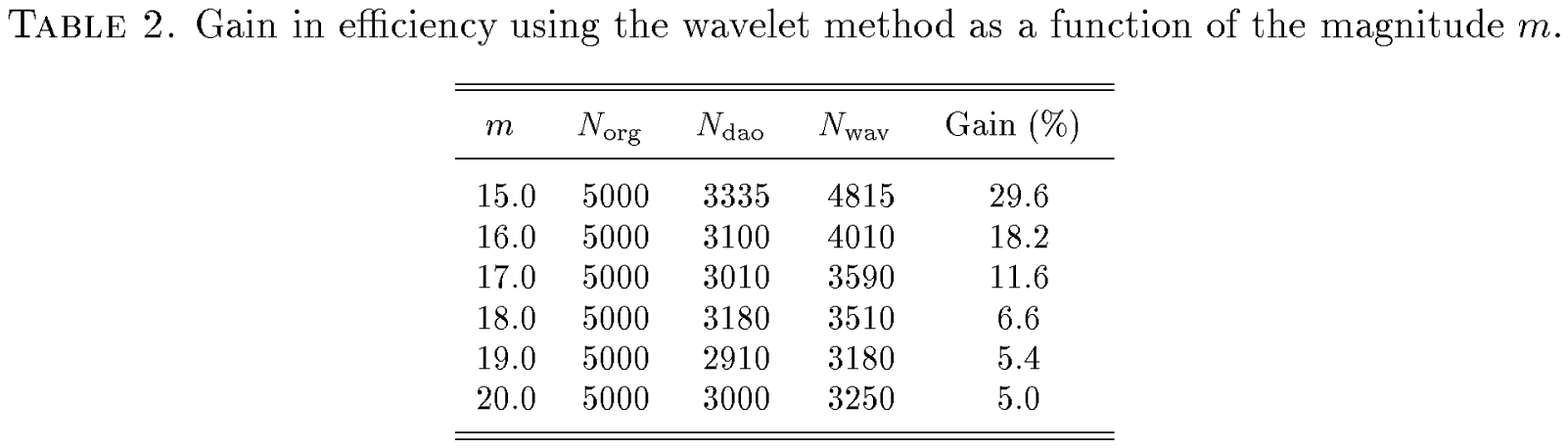}{1.5in}
{0}{90}{90}{-272}{-295}
\end{figure}

One final test was carried out in order to see how the efficiency of the
method depends on the magnitudes of the sources. We prepared several test
images with point sources of equal magnitudes: 15, 16, 17, 18, 19, and 20 mag.
The concentration parameter adopted was $c = 1.5$. Table 2 lists
the results of this experiment. In column 1, the magnitude $m$ is indicated.
The remaining columns have the same meanings as in Table 1.

\begin{figure}[t]
\figurenum{1}
\plotfiddle{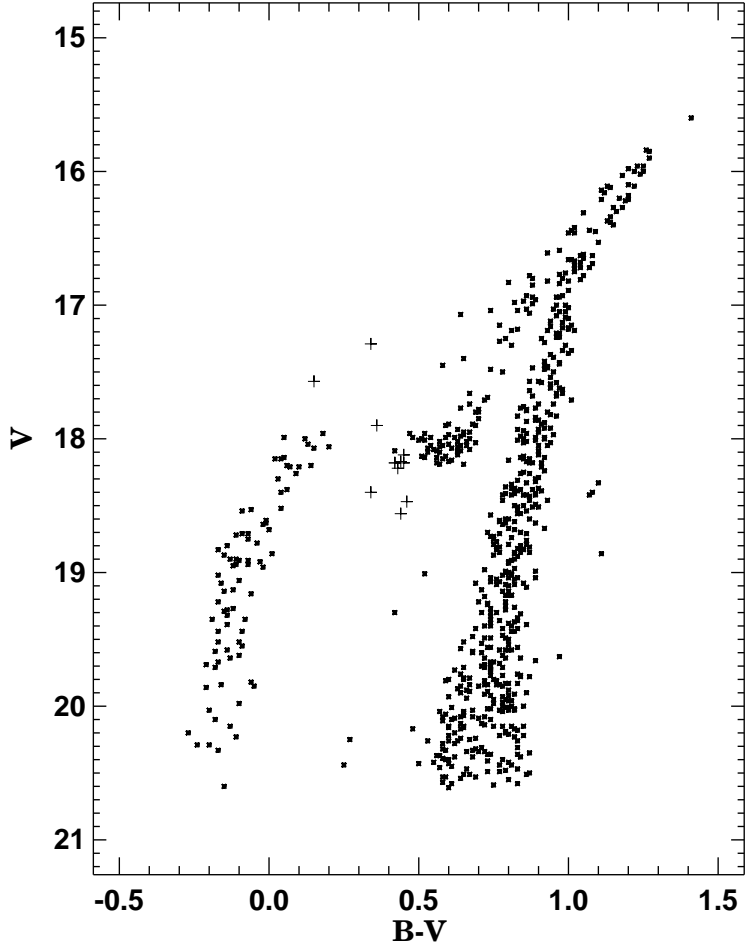}{4.825in}
{0}{90}{90}{-212.88}{-324}
\caption{The observed $V$ vs. $\bv$ CMD for the central regions of NGC 6229
 derived from this work. The new candidate RR Lyrae variable stars are shown 
 as plus signs. The 7 RR Lyrae variables which are known to be present in
 the field have not been plotted.}
\end{figure}

Based on the above tests we can conclude that the wavelet transform technique
is a rather powerful method for analyzing the photometric data in crowded 
fields.

\subsection{CMD}

The photometry of the stars on the frames was performed by means of the PSF
fitting package {\sc DAOPHOT} using a list of $X$, $Y$ coordinates determined
from the ``resolved" images.

\begin{figure}[t]
\figurenum{2}
\plotfiddle{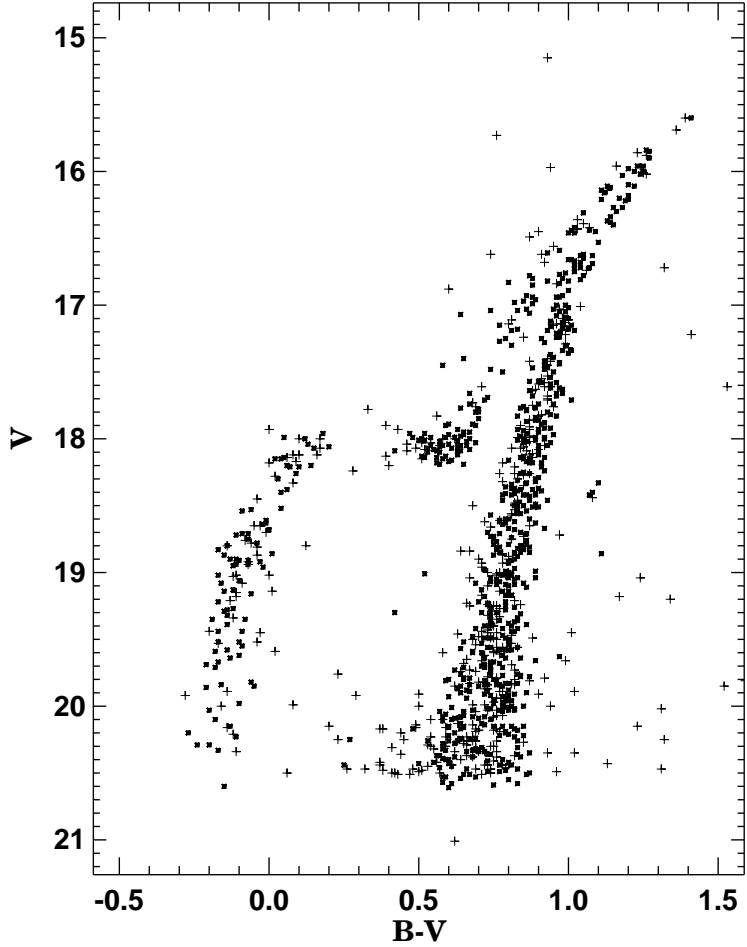}{4.825in}
{0}{90}{90}{-212.88}{-324}
\caption{The composite CMD of NGC 6229 based on our CCD data
 (\mbox {\boldmath $_{^{\times}}$} ) and the data of CFT91 (+).
 Known and candidate RR Lyrae variables have been omitted.}
\end{figure}

        The instrumental values were corrected for extinction and transformed
to the standard $BV$ system using a technique similar to that described by
Christian \& Heasley (1988). The instrumental $b$ and $v$ magnitudes
were converted to the standard $B$ and $V$ system by means of the equations:\\

\noindent
$\displaystyle
V  =  v  + 2.806 - 0.301\, X + 0.011\, (b-v)$, \hfill \\
$\bv =  -0.629 - 0.123\, X + 1.022\, (b-v)$, \hfill \\ \\
\noindent
where $X$ represents the airmass.
The coefficients were determined by reducing the frames of the standard fields
in NGC 7006 (Christian et al. 1985) using an aperture photometry algorithm.
The rms deviations of the reduced CCD photometry from the standard values are
0.012 mag in $V$ and 0.015 mag in $B$, over the range of the standards.

        Our final photometric list contains 700 stars in the
central $1.5\arcmin \times 2.0\arcmin$ field of NGC 6229.
Data are available by E-mail request to the first author.
In Fig. 1, we show the CMD obtained from the present investigation. New
candidate RR Lyrae variables are shown as plus signs (cf. Sect. 5.1).

As already stated, CFT91 investigated the more external regions
of the cluster beyond $0.8\arcmin$ from the cluster center, while our CMD
covers the core region. We have transformed our coordinates and magnitudes
to the coordinate and magnitude system of CFT91 using 60
stars in common between the two studies. We then added the
400 external stars by CFT91 to our internal list of 700 stars
in order to generate a common list of stars and produce the combined CMD
that is displayed in Fig. 2. Known or suspected RR Lyrae variables have
not been included in this plot.

\begin{figure}[t]
\figurenum{3}
\plotfiddle{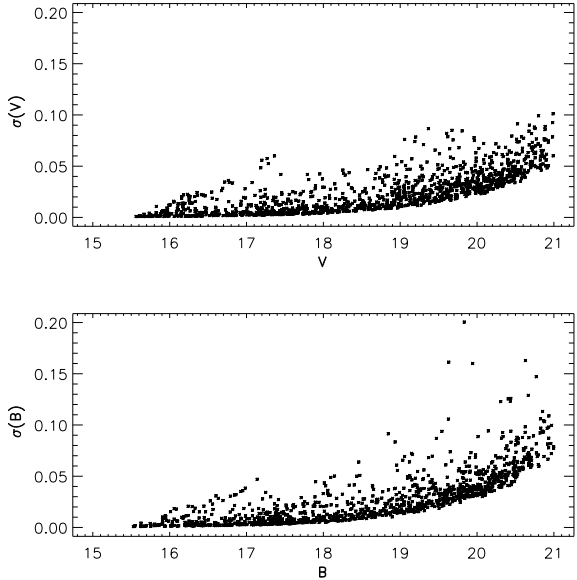}{3.00in}
{0}{55.0}{55.0}{-144.66}{-198}
\caption{Internal errors of the CCD photometry. The rms values $\sigma$ of the
 frame-to-frame scatter in the magnitudes $V$ and $B$ are plotted
 for each star as a function of $V$ (upper panel) or $B$ (lower panel).}
\end{figure}

\begin{figure}[t]
\figurenum{4}
\plotfiddle{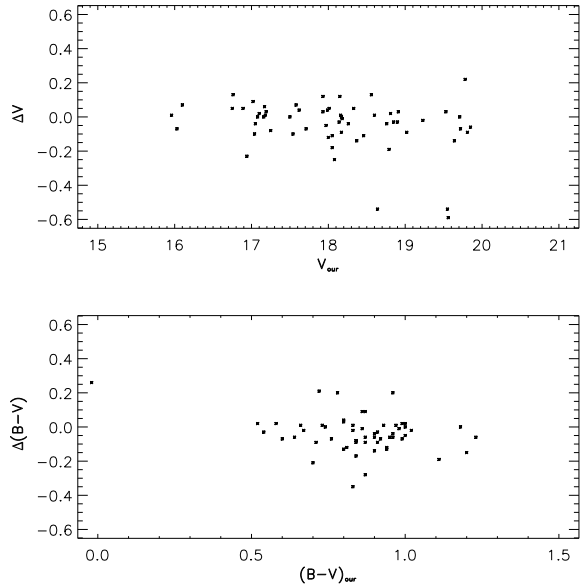}{3.00in}
{0}{55.0}{55.0}{-144.66}{-198}
\caption{ Comparison of the $V$ magnitudes (upper panel) and $\bv$ colors
 (lower panel) derived here with the photometry of CFT91. The comparison
 is in the sense of this work minus CFT91.}
\end{figure}

\section{Error analysis}

\subsection{Photometric errors}

In order to evaluate the internal accuracy of our CCD measurements, we have used
the procedure described by Buonanno et al. (1995). The instrumental magnitudes
obtained for each frame were adjusted to a common reference level and averaged
for the final calibration. The rms values were computed
according to the formula given by Buonanno et al.

        The results for each star in the $B$ and $V$ filters are plotted in
Fig. 3 as a function of the final adopted magnitude. These internal errors were
averaged over the stars in different magnitude intervals to give the following
mean errors: $\epsilon_V$ = 0.008 and $\epsilon_B$ = 0.009 for
$15.50 < V({\rm mag}) < 19.50$,
and $\epsilon_V$ = 0.03 and $\epsilon_B$ = 0.038 for
$19.50 < V({\rm mag}) < 21.00$, where the errors are expressed in magnitudes.

\subsection{Comparison with CFT91}

We have compared our data with the photometry of CFT91 in order to check for
any systematic effects. The residuals in the measurements of stars
in common between the two studies are plotted in Fig. 4 as a function of the
magnitudes and colors derived in the present investigation.
Inspection of these diagrams reveals no systematic trend in the $V$
magnitudes. There is, however, a small systematic color shift
[$\Delta (\bv ) = -0.02$], with our photometry being a little redder than the
photometry of CFT91.

\subsection{Field star contamination}

It is obviously necessary to check for field star contamination before 
discussing the detailed structure of the cluster's CMD. Based on the Bahcall \&
Soneira (1980) model of the Galaxy, Ratnatunga \& Bahcall (1985) have predicted
field star densities in the direction of NGC 6229. Table 3 gives the number
of objects expected per bin of magnitude and color in the area covered by our
observations. As one can easily see, field stars should not seriously affect
the structure of the CMD.

\begin{figure}[t]
\plotfiddle{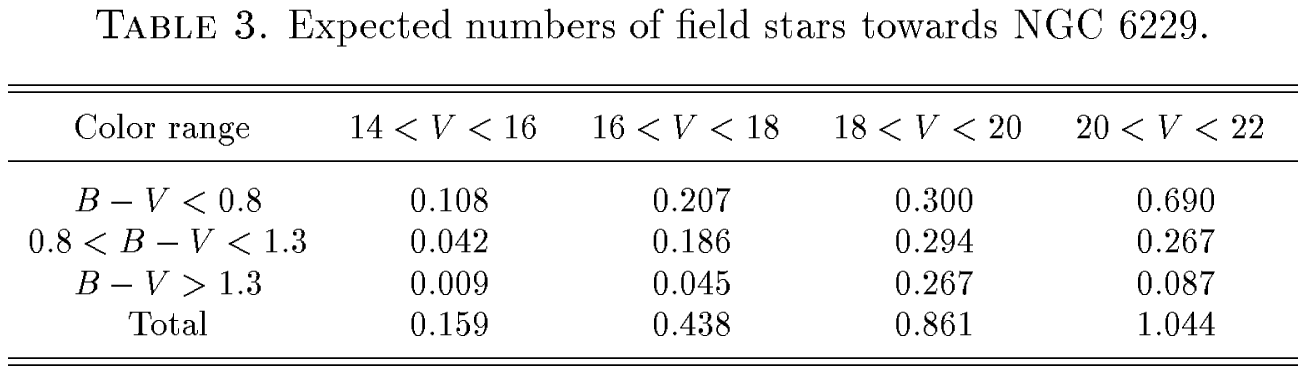}{1.5in}
{0}{90}{90}{-272}{-295}
\end{figure}

\subsection{Completeness}

        To determine the completeness function, artificial star experiments
are usually carried out (Mateo 1988; Stetson \& Harris 1988; Stetson 1991a).
In the present case, 10 different sets of 100 randomly placed stars were added
to the ``original" $B$ and $V$ frames. The frames were ``resolved" by means of 
the wavelet transform (cf. Sect. 2) and the resulting images were then 
re-reduced. Since completeness is a function of the distance from the cluster
center and of the limiting magnitude, we performed the procedure in three
radial annuli with  $ r <  0.26\arcmin $, $ 0.26\arcmin < r < 0.46\arcmin$ and
$0.46\arcmin < r < 1.13\arcmin$. Fig. 5 shows the completeness functions in the
different annuli plotted as a function of the $B$ and $V$ magnitudes.
As can be seen, the CMD becomes severely incomplete for 
$V \gtrsim 19.5$ mag for the innermost zone, $V \gtrsim 20.0$ mag for the
intermediate zone, and $V \gtrsim 20.5$ mag for the outer zone of the cluster.
For this reason, we cannot rule out the possibility
that the blue HB of NGC 6229 is actually more extended than shown in
Figs. 1 and 2. This possibility will be discussed further in Sect. 5.4.

\begin{figure}[t]
\figurenum{5}
\plotfiddle{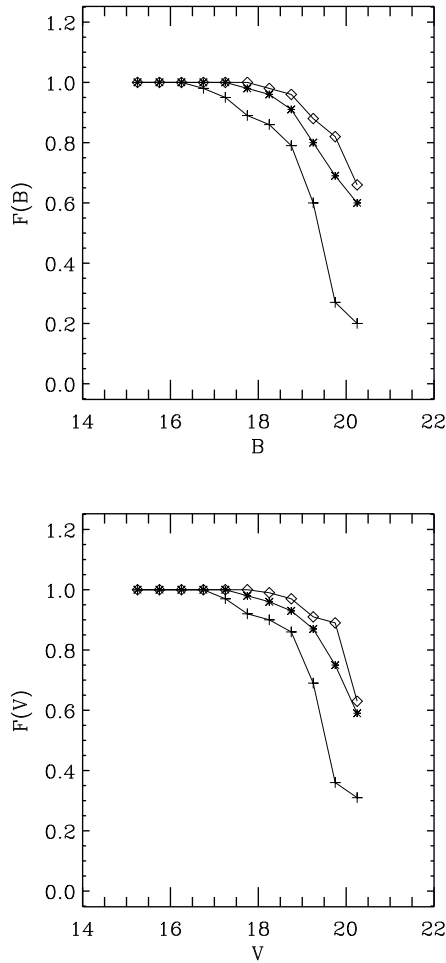}{4.825in}
{0}{90}{90}{-149.82}{-324}
\caption{Completeness curves in $B$ (upper panel) and $V$ (lower panel) 
 for annuli located at different distances from the cluster center
 (+: $r < 0.26\arcmin$; $\ast$: $0.26\arcmin < r < 0.46\arcmin$;
 $\Diamond$: $0.46\arcmin < r < 1.13\arcmin$).}
\end{figure}

We turn now to an analysis of the NGC 6229 CMD.

\section{RGB morphology}

In Fig. 6, we superimpose the M3 (NGC 5272) ridgeline (Buonanno et al. 1994)
on the CMD of NGC 6229 from Fig. 2. In producing this
plot, we have adjusted the M3 ridgeline to the NGC 6229 distance, which was 
achieved by forcing the two HBs to coincide at the RR Lyrae level
(implying a $\delta V = 2.42$ mag shift), and have also taken into account 
a small, but possible, reddening difference between the two clusters
[$\delta E(\bv )_{\rm NGC\, 6229 - M3} = 0.01$ mag].

\begin{figure}[t]
\figurenum{6}
\plotfiddle{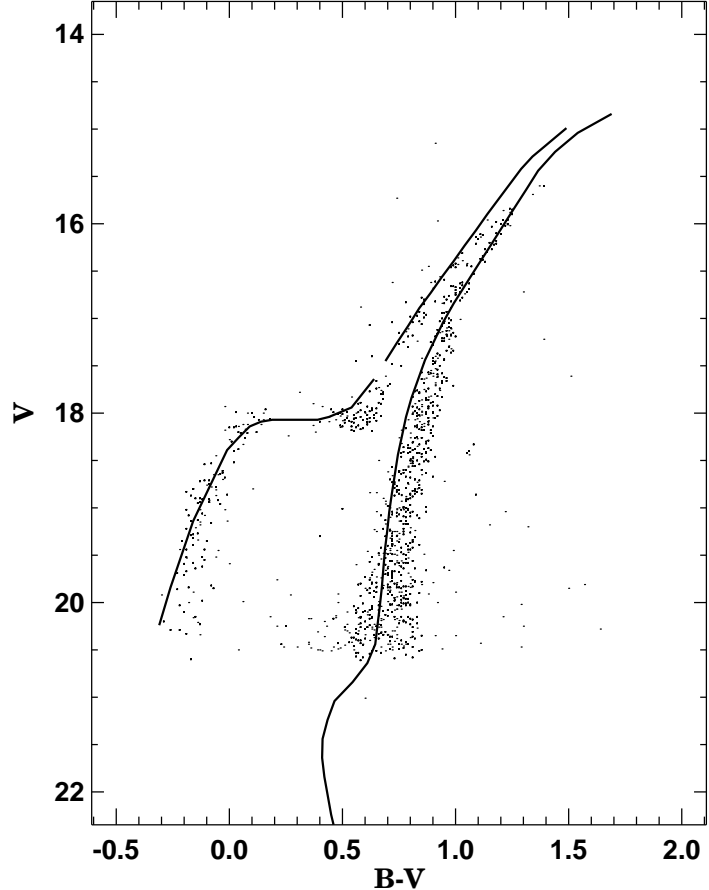}{4.5in}
{0}{85}{85}{-201.63}{-315}
\caption{Same as Fig. 2, but with the M3 ridgelines from Buonanno
 et al. (1994) superimposed on the CMD. Shifts in the latter have been applied
 both in $V$ ($\delta V = 2.42$ mag) and in $\bv$ [$\delta (\bv ) = 0.01$ mag],
 so as to account for the differences in distance modulus and reddening between
 the two clusters.}
\end{figure}

The two clusters have quite comparable metallicities, with 
${\rm [Fe/H]_{M3}} = -1.57$ (Harris 1996) or possibly slightly higher 
(Kraft et al. 1992; Carretta \& Gratton 1996; Ferraro et al. 1996).
As expected for similar metallicity clusters, the slopes of the upper RGBs 
($V \lesssim 17$ mag) are quite similar. However, in the range 
$17 \lesssim V ({\rm mag}) \lesssim 19$ there is a discrepancy, with the
M3 RGB being significantly hotter. This same discrepancy is also present
in the photometric data given in Table 3 of CFT91. We do not have a good 
explanation for this problem. If, however, combined metallicity and reddening 
effects were involved, one would then not expect the upper RGBs of the
two clusters to coincide. Another unusual feature of the NGC 6229 RGB is an 
abrupt change in slope at $(V, \bv ) \simeq (16.9, 1.0)$.

\begin{figure}[t]
\figurenum{7}
\plotfiddle{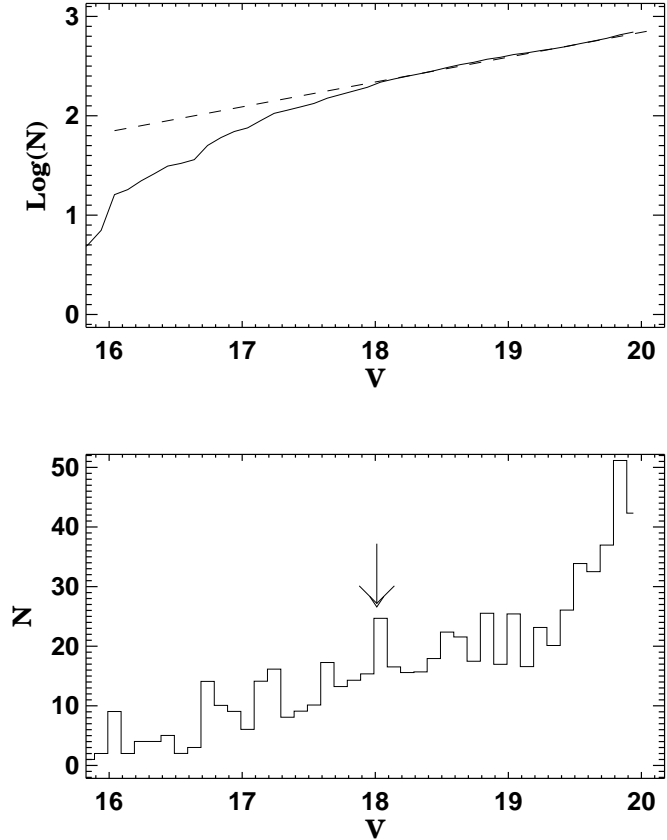}{4.25in}
{0}{80}{80}{-189.25}{-297}
\caption{The observed integrated (upper panel) and differential (lower panel)
 luminosity function for the NGC 6229 RGB.
 The arrow indicates the possible location of the RGB ``bump" in the
 cluster. These plots should be compared with the similar ones presented by
 Fusi Pecci et al. (1990) for M3.}
\end{figure}

We have attempted to apply the Fusi Pecci et al. (1990) procedure to locate
the RGB luminosity function ``bump" in NGC 6229. The purpose of their technique
is to identify the break in the slope of the cumulative RGB luminosity function
which occurs when the H-burning shell reaches the discontinuity in the H 
abundance profile that was caused by the previous deep penetration of the 
convective envelope during the first dredge up. The RGB luminosity function 
for NGC 6229, plotted in Fig. 7, tentatively suggests that this break in the 
slope coincides with a feature in the differential luminosity function,
at $V \simeq 18.05 \pm 0.07$, which resembles very closely (both in
sharpness and location) the feature in the M3 luminosity function that 
Fusi Pecci et al. have associated to the bump in that cluster (cf. their 
Figs. 3 and 4). From Eq. (7) in Fusi Pecci et al., assuming 
${\rm [Fe/H]} = -1.44$ and ${\rm [\alpha /Fe]} = 0$ for the cluster,
we would expect the bump to be present at $\approx 0.1$ mag above the
lower envelope of the HB. Since the lower envelope of the HB is at 
$V({\rm HB}) \simeq 18.20 \pm 0.03$ mag, we derive
$\Delta V ({\rm bump - HB}) \simeq -0.10 \pm 0.08$. Our results are thus 
in good agreement with those presented by Fusi Pecci et al. for their 
sample of clusters, {\em if} the noted feature proves indeed to be the 
RGB bump. We note, however, that the tentative NGC 6229 bump does not 
seem to be as sharp as those found for other clusters of similar 
metallicity (cf. Figs. 3 and 4 in Fusi Pecci et al., and Ferraro et 
al. 1996 for an updated discussion of the case of M3).

Finally, we have analyzed the width of the RGB of NGC 6229 for signs
of an internal metallicity dispersion. The ``intrinsic" width of the RGB was
estimated on the basis of the Simoda \& Tanikawa (1970) and
Da Costa \& Armandroff (1990) methods. The results are consistent with
no internal metallicity dispersion in NGC 6229.

\section{HB morphology}

\subsection{Variable stars}

        The Third Catalog of Variable Stars in Globular Clusters (Sawyer Hogg
1973) lists 22 variable stars in NGC 6229. One of them (V8) is a Population II
cepheid, another (V22) is probably a slow variable star, and
the remaining objects are RR Lyrae-type variables. It is
instructive to note that the presence of Population II cepheids is
known to be restricted to GCs containing well-developed blue HBs
only (Wallerstein 1970; Smith \& Wehlau 1985).
For this reason, the existence of
even a single such star in NGC 6229 could have been used as a strong
argument against the results of Cohen (1986), who had found the HB of NGC 6229
to be extremely red.

        In their survey CFT91 have obtained the magnitudes and colors for
20 variable stars from Sawyer Hogg's (1973) catalog (cf. Table 2 in CFT91) and
have found 1 possible variable star (their No. 155).
       On our frames only 8 variable stars from the Sawyer Hogg catalog
were actually detected because of our small field of view: V5, V6, V8, V11,
V12, V15, V16, and V20.

        In order to search for as yet undetected variables, the method proposed
by Kadla \& Gerashchenko (1982) was used. Such a method is based on the
determination of the positions of the RR Lyrae variables on a CMD obtained 
from measurements of two images taken within a time interval that is much 
shorter than the periods of the variables. 
Kadla \& Gerashchenko's original application of the method was devoted to
the GC M3, where the instability strip region is much more heavily populated 
than in NGC 6229. After obtaining $B$ and $V$ frames with a separation in time
of only 10 min, they showed that the RR Lyrae stars occupied a quite 
well-defined region in the CMD (cf. their Fig. 1), with only minor overlap with 
the stable red and blue HB. New variable star candidates could then be detected,
even without several different frames being available for the cluster, on the
basis of their ``instantaneous" positions on the CMD. It should be emphasized 
that, for the method to work properly, the $B$ and $V$ magnitudes {\em must} be 
measured at the same phase. This will, of course, ensure that the large 
majority of the RR Lyrae stars in such a CMD will be located within the 
RR Lyrae ``gap."

In the case of NGC 6229, as already emphasized, not many known RR Lyrae 
variables are present in our observed field. However, as shown in Fig. 1, most 
of the 12 new RR Lyrae candidates obtained with the method stand out quite 
clearly in the CMD and are unlikely to be misidentifications. A second pair
of $B$ and $V$ images that we have obtained to check this result again shows 
all of these 12 candidates to lie within the instability strip region.
The star lying at $(V, \bv ) \simeq (18.1, 0.42)$ has uncertain photometry
in one of the two pairs of $B$, $V$ images. From 
inspection of this plot and of Fig. 1 in Kadla \& Gerashchenko (1982), 
we deem it highly unlikely that a significant number of RR Lyrae 
variable candidates has been missed in the present study. 

Information 
about the new RR Lyrae candidates is given in Table 4. In columns 2 and 3 
of this table, we list the coordinates (in $\arcsec$) of the suspected variables
in the Sawyer Hogg (1973) system. The next two columns contain $V$ and $\bv$ 
values, determined by only one image pair. $R_{\rm c}$ is the projected radial 
distance (in $\arcmin$) from the cluster center. We are currently undertaking
a new program to check the variability of these objects and obtain detailed 
light curves for the confirmed RR Lyraes.

\begin{figure}[t]
\plotfiddle{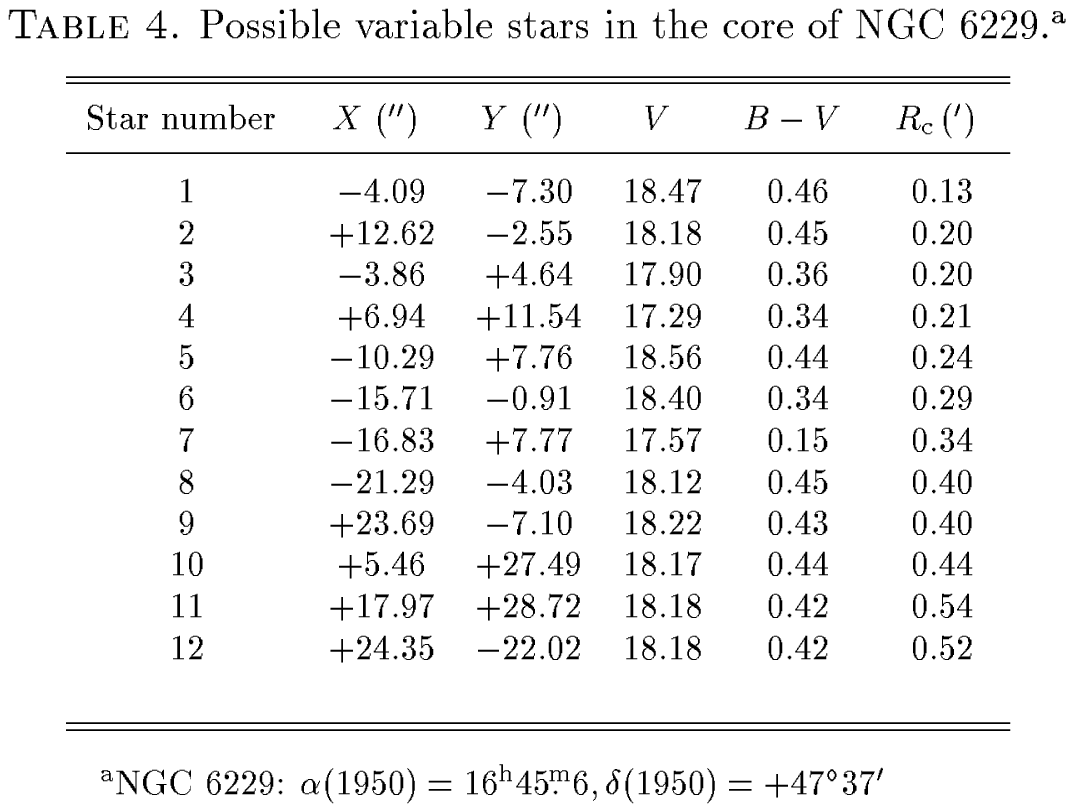}{3.25in}
{0}{90}{90}{-272}{-223}
\end{figure}

\subsection{Number counts}

        The HB of NGC 6229 reveals many interesting features. We confirm the
finding by CFT91 that the HB morphology of NGC 6229 is peculiar and atypical
for an outer-halo cluster. In particular, NGC 6229 displays a well-populated
blue HB tail, extending to quite blue colors and to a faint luminosity level
($\simeq 2.5$ mag below the RR Lyrae level in $V$). The presence of such an
extended blue HB tail is a very unusual property for an outer-halo GC,
although it is consistent with the presence of a Population
II cepheid in this cluster (cf. Sect. 5.1).

        Using the combined list of stars from this work and from the
CFT91 study, the number of detected HB stars has now been increased to 235,
or 266 after applying a completeness correction.
This is to be compared with the 87 HB stars previously studied by CFT91.
Our study thus improves
considerably the degree of statistical significance of the detected HB
features.

        Following Fusi Pecci et al. (1993), we have estimated the degree of
``statistical significance" of the present sample of HB stars. We have
calculated their parameter
${\rm Frac}_{\rm HB} = N^{\rm obs}_{\rm HB}/N^{\rm tot}_{\rm HB}$, where
$N^{\rm obs}_{\rm HB}$ and $N^{\rm tot}_{\rm HB}$ are the observed and total
number of HB stars, respectively, within the GC. The total number of stars
expected to populate the HB of NGC 6229 is $\approx 250$, according to Eq.
(14) of Renzini \& Buzzoni (1986), assuming $\log L_{\rm tot} \simeq 5.1$
[as obtained from the integrated $V$ magnitude $M_V$ listed by Harris (1996), 
and assuming $M_V \approx M_{\rm bol}$ (Fusi Pecci et al. 1993)]. The implied 
degree of statistical significance of the present sample is thus quite high, 
although it should be noted that the total HB population cannot be accurately 
predicted in this way due to the uncertainties in, e.g., the ``evolutionary 
flux," the cluster distance modulus, and the bolometric correction.

To determine the number of stars in the HB, RGB, and
asymptotic giant branch (AGB) phases,
we followed the procedure used by Buzzoni et al. (1983). The data
were corrected for completeness and the estimated numbers rounded to the closest
integer value. The results for the number of stars on the different
branches of NGC 6229 are presented in Table 5. In this table, $B_W$ represents 
the number of HB stars in the color range $-0.02< (\bv )_0 <0.18$, and $B2$ 
the number of HB stars bluer than $(\bv )_0 = -0.02$ (cf. Buonanno 1993). 
The data quoted from CFT91 do
not include the 60 stars in common with our study. We have thus counted up
separately the stars with  $r_{\rm c} < 1.5 \arcmin$ from our data and with
$r_{\rm c} > 1.5 \arcmin$ from the CFT91 data. We point out the
uncertain photometry and the small number of measured stars within the central
$15 \arcsec$.

\begin{figure}[t]
\plotfiddle{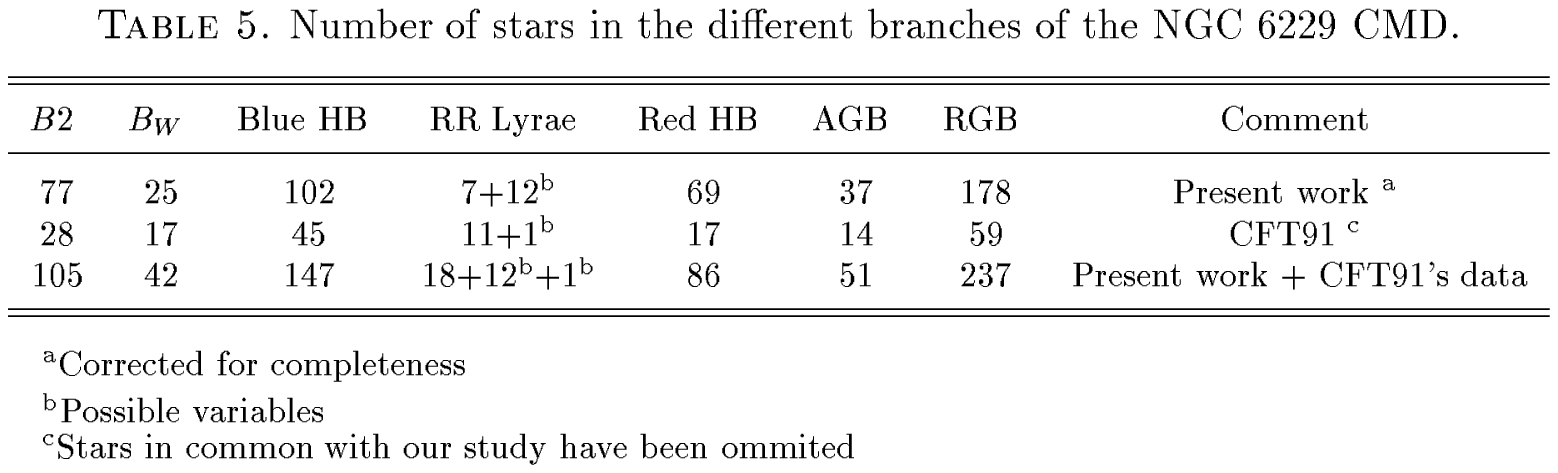}{1.5in}
{0}{90}{90}{-272}{-295}
\end{figure}

Such number counts may be employed to estimate several useful HB
morphology parameters. One of the most traditional is the so-called
Dickens (1972) type ($DT$). Harris (1996) has estimated a $DT = 3$ for NGC 6229,
based on the CFT91 CMD. From Fig. 5 in Dickens' original paper, however, one
clearly sees that such an HB morphology classification should correspond to a
high density of stars within the instability strip, and to a significantly
smaller number of stars on the blue HB and (especially) the red HB. However,
the number counts for NGC 6229 show a relative deficit of RR Lyrae variables
compared to the total HB population. From Table 5, one finds
$B:V:R = 0.59:0.08:0.33$ (where $B$, $V$, and $R$ stand for the relative
number of blue HB stars, RR Lyrae variables, and red HB stars, respectively).
The corresponding number ratios for the typical $DT = 3$ cluster M5 (NGC 5904)
are $B:V:R = 0.56:0.24:0.20$, according to the compilation presented by Lee
et al. (1994). For comparison,
M3 (NGC 5272) has a $DT = 4$ HB with $B:V:R = 0.33:0.42:0.25$, and
M92 (NGC 6341) has a quite blue $DT = 2$ HB, with $B:V:R = 0.90:0.08:0.02$.
We thus conclude that the classification of NGC 6229
as a $DT \approx 3$ cluster is probably not correct.

Following Harris' (1996) classification scheme, and using the number counts
in Table 5, we believe it would probably be more
appropriate to identify this cluster as a $DT = (1\pm 1)/(7\pm 1$)
cluster---i.e., a cluster with a bimodal HB distribution, resembling those
of  NGC 2808 (Ferraro et al. 1990: $DT = 0/7$, $B:V:R = 0.23:0.01:0.76$)
or NGC 1851 (Walker 1992: $DT = 1/6$, $B:V:R = 0.28:0.11:0.61$), but with a
larger relative population of blue HB stars than in these two other bimodal-HB
clusters. Other good, but perhaps less well-known, bimodal-HB candidates 
are M4 (NGC 6121), with $B:V:R = 0.34: 0.25: 0.41$; 
NGC 6362, with $B:V:R = 0.19: 0.04: 0.77$; NGC 6723, with
$B:V:R = 0.37: 0.19: 0.44$; and M75 (NGC 6864), with $B:V:R = 0.27: 0.07: 0.66$
(number counts are from Lee et al. 1994). None of these has been
classified as a bimodal-HB cluster by Harris. A bimodal HB has been assigned
to M62 (NGC 6266) by the latter author. However, Brocato et al. (1996) have 
very recently derived $B:V:R = 0.58: 0.26: 0.16$ for M62, thus ruling out a 
bimodal HB for the cluster. We will return to this question of bimodality 
in Sect. 5.3.

Table 6 lists a number of other parameters obtained from our number counts
for NGC 6229. We include in this table the inferred values for
several of the new HB morphology indicators that Fusi Pecci et al. (1993) and
Buonanno (1993) have recently defined.
The quantities listed in Table 6 are as follows: $N_{\rm HB}$ is the
number of HB stars, $N_{\rm RGB}$ is the number of RGB stars brighter than the
HB level, $N_{\rm AGB}$ is the number of AGB stars,
$(\bv )_W$ is the derived mean color for stars falling
in the range $-0.02< (B-V)_0 <0.18$,
and $B_W = N^{\rm HB}_{-0.02 < (B-V)_0 < 0.18}$
is the number of stars in this color interval.
The values in parentheses are for the case when the suspected variable
stars are included in the calculations. The errors were estimated on the basis
of Poisson statistics.

\begin{figure}[t]
\plotfiddle{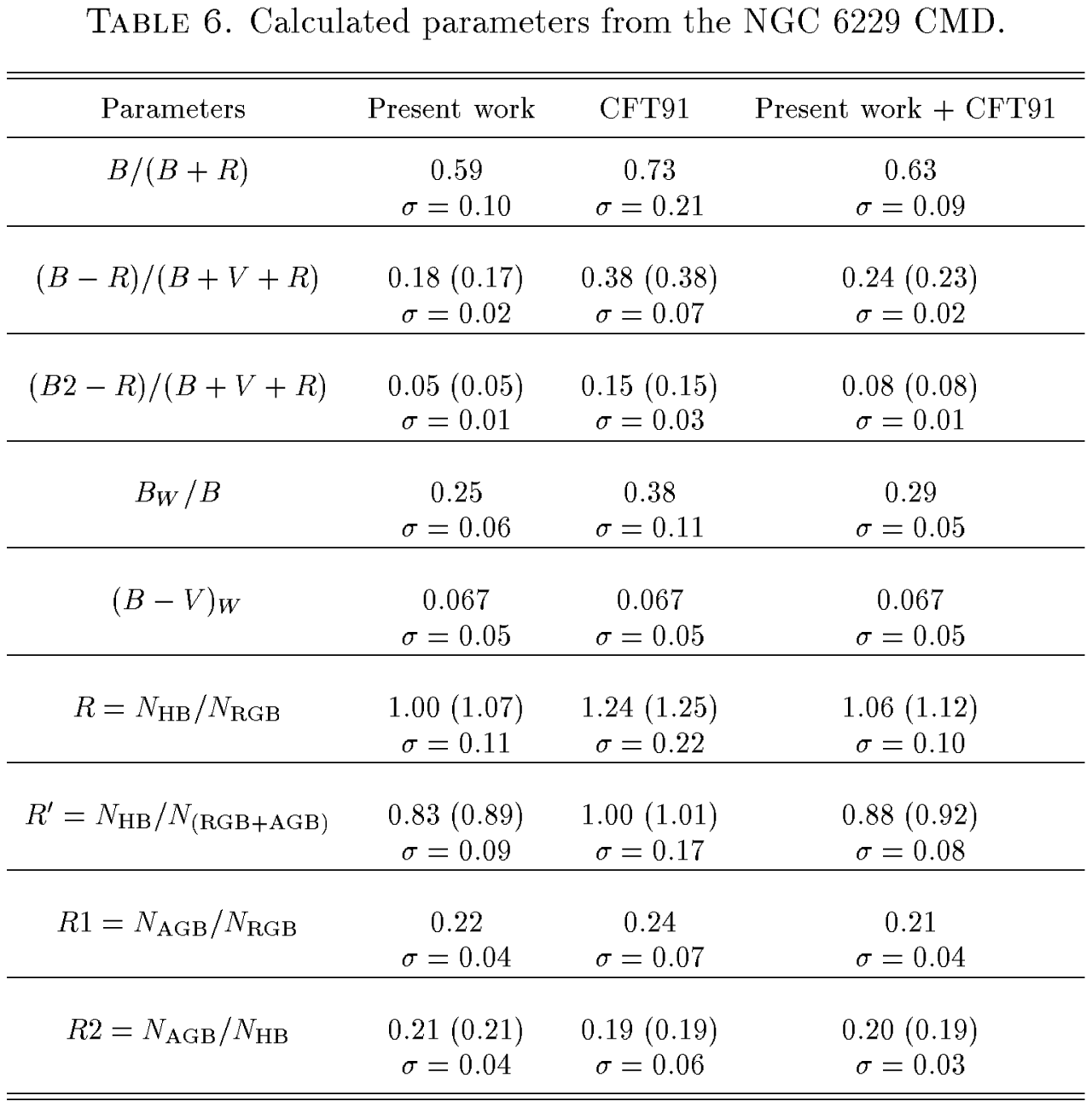}{5.00in}
{0}{90}{90}{-272}{-151}
\end{figure}

The values derived for $R$ of $1.00 \pm 0.11$ (using our data only) and
$1.06\pm 0.10$ (from all data) and $R'$ of $0.83 \pm 0.09$ and $0.88 \pm 0.08$,
respectively, are lower than average (Buzzoni et al. 1983).
These parameters are sensitive to the envelope helium abundance in evolved GC
stars, and their low values in NGC 6229 are suggestive of a lower-than-average
envelope helium abundance [$Y\simeq 0.18$, according to Eqs. (11) and (12) in
Buzzoni et al. (1983); see also CFT91], although alternative interpretations
are possible (cf. Sweigart 1996).
The suggestion by CFT91 that the apparently low helium abundance of NGC 6229
may be responsible for its relatively blue HB is not supported by HB
evolutionary models, which instead predict that the HB morphology should
get {\em redder} with decreasing $Y$ (e.g., Sweigart 1987; Catelan 1993).

The values of $R$ and $R'$ would be affected by any incompleteness in the
relevant number counts. However, for the $R$ value in NGC 6229 to match the
average value of 1.41 found by Buzzoni et al. (1983), we would have to have
missed 71 HB stars or counted 50 RGB stars too many---requiring changes of
$\approx 30\%$ and $\approx 24\%$, respectively, in the corresponding number
counts. In
the case of $R'$, the corresponding changes are $\approx 33\%$ and
$\approx 25\%$, respectively.
However, as can be seen from Fig. 5, we have complete samples for AGB and RGB
stars, and a completeness correction has been applied for the blue HB stars
brighter than our photometric limit of $V \simeq 20.5$ mag. It is entirely
possible that the blue HB tail in NGC 6229 extends to fainter magnitudes, 
in which case our number counts could be missing a significant number of very 
blue HB stars (see Sect. 5.4).

\begin{figure}[t]
\figurenum{8}
\plotfiddle{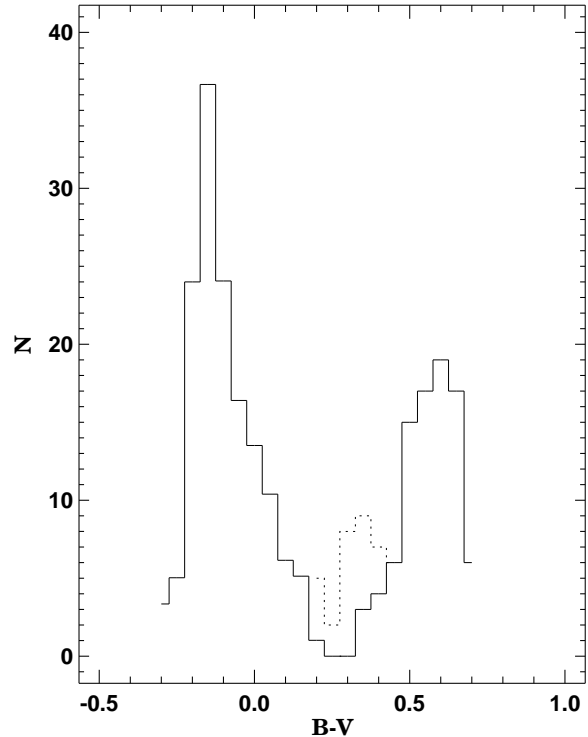}{3.75in}
{0}{70}{70}{-166.26}{-253}
\caption{The histogram showing the color distribution of HB stars.
 The approximate distribution of the RR Lyrae stars is shown
 as dashed lines. }
\end{figure}

The ratios $R1$ and $R2$ have been used to confirm the existence of
semiconvective zones outside the helium-burning convective cores of HB stars
(Renzini \& Fusi Pecci 1988).
The $R1$ ratio for NGC 6229 is similar to the one found by
Buzzoni et al. (1983) for their sample of clusters
($\langle R1 \rangle = 0.21 \pm 0.03$), while the $R2$ ratio
seems somewhat higher than average ($\langle R2 \rangle = 0.15 \pm 0.03$;
Buzzoni et al. 1983). This result might be another indication that faint blue
HB stars ($V \gtrsim 20.5$ mag) have been missed in our number counts.

\subsection{Bimodality}

There is a clear indication of a bimodal distribution in color along the HB of
NGC 6229. A well-populated red HB and an extended blue HB tail
are clearly present, but the instability strip level seems to be more scarcely
populated. The Sawyer Hogg (1973) catalog lists 20 RR Lyrae stars. To this
number should be added the one possible RR Lyrae star found by CFT91 and the
12 new possible RR Lyrae stars found in the present study
within the central field of the cluster. In order to estimate how many RR Lyrae
stars may have been missed in our observational survey, we have carried out
some numerical experiments. Our frames were taken ``simultaneously" in $B$ and
$V$, thus ensuring that the instantaneous colors of the vast majority of the 
RR Lyrae stars would place them in the RR Lyrae gap (Kadla \& Gerashchenko 
1982; cf. Sect. 5.1). As in the completeness analysis (cf. Sect. 3.4),
30 artificial RR Lyrae stars with $V$ magnitudes between 18.20 and
17.80 and $B$ magnitudes between 18.25 and  18.65 were randomly added on the
original images. The frames were then re-reduced as described in Sects. 2 and 3.
The mean recovery efficiency was $87 \%$, and the maximum number of ``missing" 
RR Lyrae stars, in such experiments, was 7 (implying a recovery efficiency of 
$\simeq 77\%$).

The histogram showing the color distribution of
HB stars is given in Fig. 8. The approximate distribution of the
RR Lyrae stars is shown with dashed lines and corresponds to 33 RR Lyrae stars
with ``mean" colors
randomly chosen so as to place them within the instability strip.
Since NGC 6229 has been classified as an Oosterhoff type I cluster (Castellani
\& Quarta 1987), from the corresponding properties of fundamental and
first-overtone pulsators we estimate that the 12 possible new RR Lyrae
stars should be divided into $\approx 3$ RRc and $\approx 9$ RRab stars.
Further photometry based on more images is necessary to obtain modern light
curves for the RR Lyrae stars, confirm the nature of the newly identified
RR Lyrae candidates, and check for the possible presence of any as-yet 
unidentified variables (cf. Sect. 5.1).

It is also worth mentioning that the red extreme of NGC 6229's blue HB seems
to be slightly brighter than the main part of the red HB of the cluster 
(cf. Figs. 1 and 2). The comparison with the more horizontal HB of M3 
(cf. Fig. 6) makes this point especially clear. Stetson et al. (1996) have 
suggested that a similar effect may be present in the bimodal-HB clusters 
NGC 1851 and NGC 2808, with potentially important implications for age 
determinations using the ``$\Delta V$" method [cf. Eq. (1) in Catelan \& de 
Freitas Pacheco (1995) and the accompanying discussion], and hence for our 
understanding of the second-parameter phenomenon. We will discuss these points 
more thoroughly in our subsequent paper (Catelan et al. 1996).

\subsection{Blue tail and ``gaps"}

CFT91 noted the possible presence of two gaps in the blue part
of the HB, at $V \simeq 18.4$ mag and $V \simeq 19.7$ mag.
The existence of the first gap is quite evident in the
CMD (cf. Figs. 1 and 2). The reality of the second gap remains more
elusive. The luminosity function of the blue HB stars gives support to
the existence of two gaps, but more
data extending to fainter magnitudes are clearly needed to settle the issue.

\begin{figure}[t]
\figurenum{9}
\plotfiddle{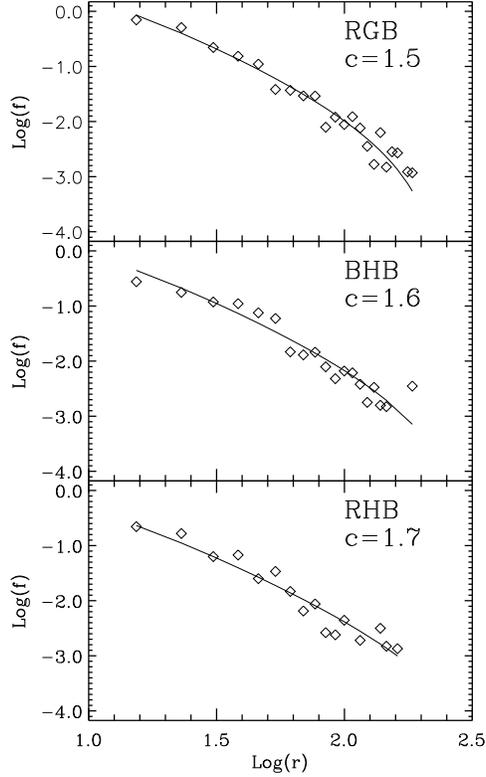}{4.0in}
{0}{85}{85}{-124.38}{-350}
\caption{The observed radial distribution of RGB stars (upper panel), blue HB 
 stars (middle panel), and red HB stars (lower panel). The best fitting
 King model with the indicated concentration parameter $c$ is plotted for
 each distribution. ``$f$" is the number of stars per squared radial bins, 
 and the values are normalized to the maxima of the corresponding
 distributions. The distance $r$ from the cluster center is given
 in $\arcsec$. In these plots, the innermost points were
 omitted, because of poor statistics.}
\end{figure}

It should be remarked that the presence (or otherwise)
of blue-HB stars fainter than $V \simeq 20.5$ mag in NGC 6229 remains quite
uncertain. In the bimodal-HB cluster NGC 2808, for instance,
Djorgovski et al. (1996)
have recently found the blue HB tail to extend to magnitudes fainter than the
main-sequence turnoff. If similar stars were present in NGC 6229, we would
not have been able to detect them directly. However, the existence of a NGC
2808--like extended blue HB tail in NGC 6229 would increase the derived
number ratios $R$ and $R'$ and decrease the ratio $R2$, compared to the
values in Table 6, while keeping the 
ratio $R1$ unchanged. This is just the sense of the difference
between our derived number ratios for NGC 6229 and the corresponding values
for other clusters. An interesting analogy is provided by the core-concentrated
cluster M80 (NGC 6093) (cf. Desidera 1996). Although the CFT91 photometry has 
a fainter threshold than the present one, it can probably still not rule out 
the presence of an extended blue HB tail. We recall that
Ferraro et al. (1990, especially their Sect. 3.1) were not able to find good
evidence for the surprisingly long NGC 2808 blue HB tail that has only very
recently been detected by Djorgovski et al. from deep {\em HST} images of the
cluster.

We conclude that deeper, high-quality photometry for NGC 6229 is urgently
needed to show whether a very long, NGC 2808--like blue HB tail is present
or not.

\section{Radial distribution}

      The center of NGC 6229 was determined following the procedure outlined by
Mateo et al. (1986). According to this technique, the cluster center is located
at $X=207.5\,\pm1.3$ px and $Y=84.9\,\pm1.9$ px.

\begin{figure}[t]
\figurenum{10}
\plotfiddle{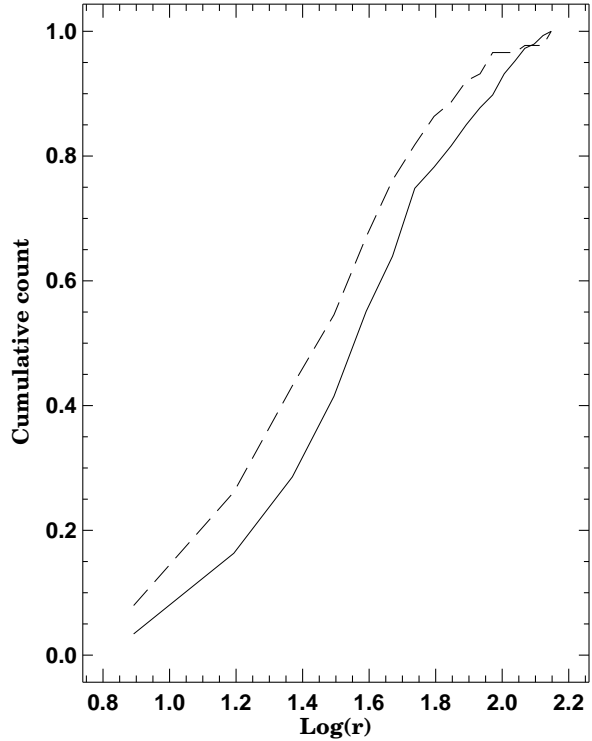}{3.75in}
{0}{70}{70}{-165.21}{-257.5}
\caption{Radial cumulative distribution of blue HB stars
 (solid line) and red HB stars (dashed line) as a function of the
 projected distance from the cluster center (given in $\arcsec$). }
\end{figure}

The structure of NGC 6229 was studied following King's (1966) iterative method
by counting stars in concentric rings of increasing radii on the $V$ image.
The data were corrected with regard to the geometrical completeness of the
annuli used. The standard $\chi^{2}$ minimization procedures were used to fit
the theoretical curves representing self-consistent King's dynamical models of
star clusters. The best fit is consistent with $r_{\rm c}=(7.9\pm 2.3)\arcsec$
and $c=1.60$. These values are in excellent agreement with those presented by
Trager et al. (1993).

In order to investigate the presence of radial population gradients,
number counts were separately made for different components of the CMD (RGB,
blue HB, and red HB) as a function of the radial distance from the cluster
center. Different sets of theoretical King (1966) models were then computed
for different values of the concentration parameter $c$. For this purpose, the
observed surface brightness profiles were derived in concentric rings of 
increasing radii on the $V$ image. A Kolmogorov-Smirnov test was applied 
to infer the best fit between theoretical and observed surface brightness 
profiles. The results are shown in Fig. 9, where $c$ is the best-fit King 
concentration parameter for the corresponding subsample. As can be seen, the 
red HB stars are more strongly concentrated toward the cluster center, in 
comparison with the blue HB stars and the RGB population.

A second approach for investigating the radial population gradients involved a
comparison of the cumulative radial distributions of stars in the blue and red
portions of the HB. A Kolmogorov-Smirnov test applied to these two radial 
distributions, which are plotted in Fig. 10, yields a probability of 94.5\% 
that they are different, again suggesting that the red HB stars 
are significantly more centrally concentrated than the blue HB population.

Following the method described by Buonanno et al. (1991), we have divided the HB
stars in two samples with equal number of stars---one including the stars 
lying inside the circle with radius $r = 0.59\arcmin$ and another including
the stars lying outside this circle. We have then calculated the HB morphology
index $(B-R)/(B+V+R)$ for these two regions separately. The resulting values
are $(B-R)/(B+V+R) = 0.16\pm 0.02$ for the inner region and 
$(B-R)/(B+V+R) = 0.31\pm 0.02$ for the outer one.
In other words, the mean color of the HB in the inner region appears
to be shifted towards the red, compared to the outer region.

These radial trends in HB morphology, if
real, are at least in qualitative agreement with those previously found for
${\rm M15 = NGC\, 7078}$
(Stetson 1991b) and NGC 6752 (Buonanno et al. 1986). We caution, however,
that deeper photometry is needed to detect any possible missing faint blue HB
stars, especially near the cluster center. For additional information on radial
population gradients in GCs, we refer the reader to the excellent discussions
presented by Buonanno et al. (1991), Djorgovski et al. (1991),
Djorgovski \& Piotto (1992), and Stetson (1991b).

\section{Concluding remarks}

Since the work by Sandage \& Wildey (1967) on NGC 7006 and the analysis of the
second-parameter phenomenon by Searle \& Zinn (1978), it has become widely
appreciated that clusters in the outer Galactic halo possess
predominantly red HB morphologies. This has frequently been interpreted as
a clue that the outer halo of the Galaxy may have taken longer to form than
the inner halo (cf. Zinn 1993 for a recent discussion). The definition of
``outer halo" has generally been based on studies of the spatial
distribution of field RR Lyrae stars and halo GCs (e.g., Saha
1985; Zinn 1985; Carney et al. 1990). In particular, since the mass
density--galactocentric distance relation for Galactic GCs presents a clear
break at $R_{\rm GC} \simeq 20\, {\rm kpc}$ (cf. Fig. 10 in Carney
et al. 1990), a dividing line between ``regular" and ``outer-halo" GCs which
is independent of the HB morphology distribution of Galactic GCs has
commonly been placed in the vicinity of 20 kpc (e.g., Ortolani 1987; CFT91),
although different dividing lines which are not independent of HB morphology
considerations have on occasion been used. In view of 
these arguments, and for consistency with CFT91, we will assume here, 
unless otherwise stated, that the Galactic outer halo begins at 
$R_{\rm GC} \simeq 24\, {\rm kpc}$.

\begin{figure}[t]
\plotfiddle{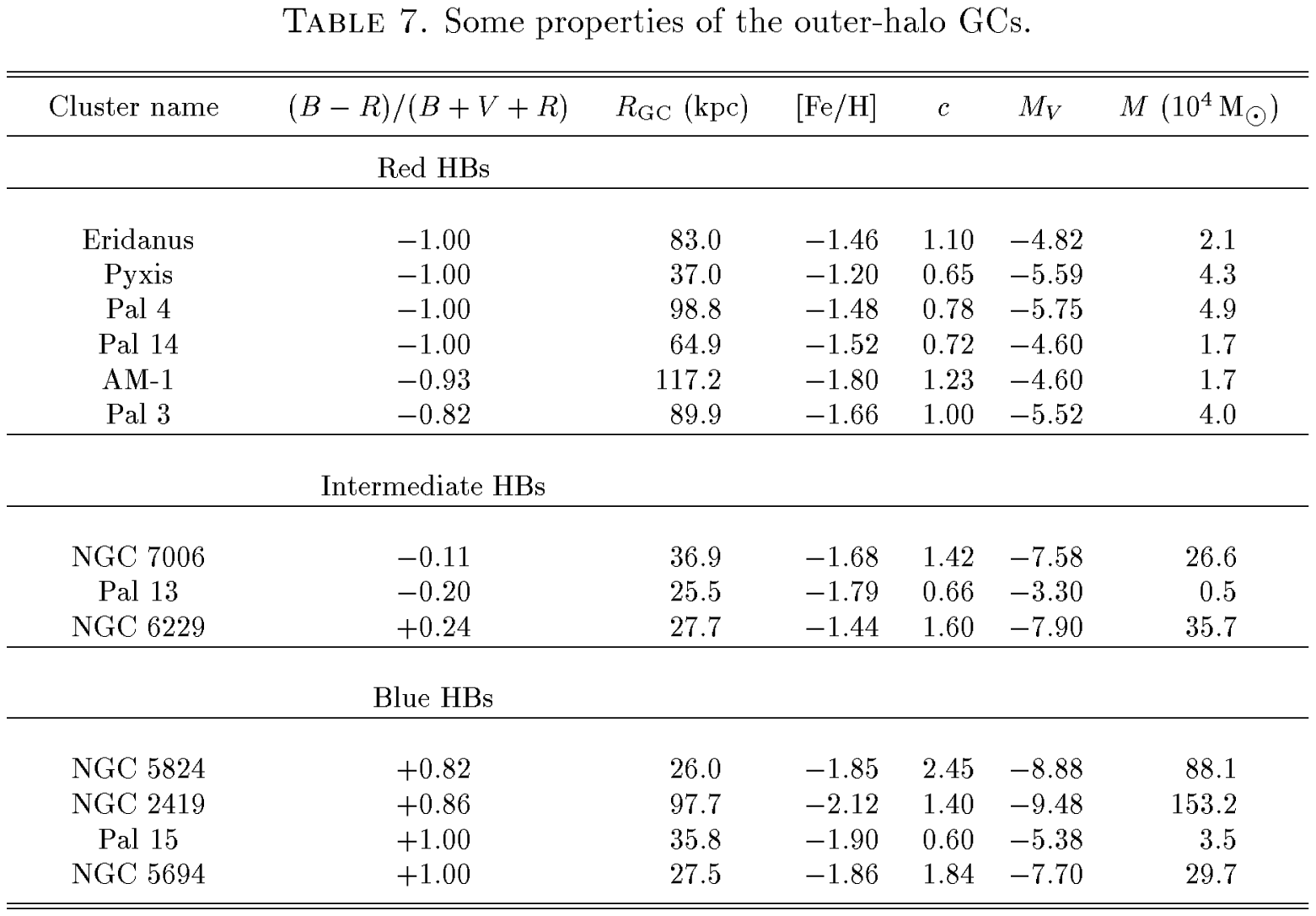}{3.75in}
{0}{90}{90}{-272}{-205}
\end{figure}

Reasonably accurate CMD data are now available for some 13 outer-halo
globulars. In Table 7, we summarize our present knowledge of the HB morphology
for these outer-halo objects. The listed data [which also include the 
corresponding $R_{\rm GC}$, [Fe/H], $c$, $M_V$, and total estimated
mass $M$ of the cluster (obtained from $M_V$ as in Djorgovski 1993)] 
have been obtained from the catalog
by Harris (1996), to which we address the reader for the detailed references.
For NGC 7006, we adopted the HB morphology provided by Buonanno et al. (1991).
The outer-halo globular AM-4 ($R_{\rm GC} \simeq 24.2$ kpc, 
${\rm [Fe/H]} \simeq -2.0$, $M_V \simeq -1.5$; Harris 1996) has not been
included due to the fact that it entirely lacks stars brighter than about
the main-sequence turnoff (Inman \& Carney 1987).
From inspection of this table, one easily concludes that the standard view that
outer-halo GCs possess predominantly red HB types is not
valid. In fact, the number of outer-halo clusters with very blue HBs (4)
is quite comparable to the number of outer-halo clusters with very red HBs
(6). If one assumes that clusters with $20 \leq R_{\rm GC}({\rm kpc}) \leq 24$
are also outer-halo members, one may then even conclude that the outer halo
has a predominantly {\em blue} HB (cf. Fig. 11): in this galactocentric
distance range, all clusters listed by Harris (Arp 2, NGC 4147, NGC 5634,
and NGC 7492) have blue HB types. In addition,
one should be aware that, in the mean, the outer-halo clusters with bluer HB
types seem to be significantly more luminous, and thus presumably also more
massive, than those with redder HB types (Fig. 11d). From the masses 
presented in Table 7, one finds that the total mass contained in the six 
outer-halo clusters with red HBs is about {\em fifteen} times smaller than
the total mass contained in the four outer-halo globulars with blue HBs, 
and even about 3.4 times smaller than the mass in the form of intermediate-HB 
clusters. Thus, if these GCs are representative survivors of the larger 
protogalactic fragments that may have led to the formation of the Galaxy 
(Searle \& Zinn 1978; Zinn 1993), it is but reasonable to conclude that a 
more substantial fraction of the mass contributed by these fragments to the 
outer Galactic stellar halo was associated with stellar populations with 
predominantly {\em blue}, and {\em not} red, HB types---contrary to the usual 
assumption. In fact, if the outer-halo field population originated 
from dissolved GCs similar to the ones currently found there, its HB 
morphology would clearly be very blue. 

We do confirm the trend discussed by CFT91 that the
outer-halo clusters with redder HB types tend to have higher
metallicities and be located farther away from the Galactic center, in
comparison with the outer-halo clusters with bluer HBs (cf. Figs. 11a and 11b).
In particular, outside $R_{\rm GC} = 40$ kpc, five out of six clusters seem
to possess red HB types. However, the cluster with a blue HB that is found in 
this region, NGC 2419, is {\em much} more massive than any of the red-HB 
clusters lying at comparable galactocentric distances (cf. Fig. 11d and 
Table 7). Racine \& Harris (1975) have argued that NGC 2419 may have a 
relatively small ($\simeq 24$
kpc) perigalactic distance, and currently be near its apogalacticon. If this
cluster formed near its perigalacticon---a possibility that Racine \& Harris
advocate---its peculiar position among the remainder of the outer-halo
globulars with $R_{\rm GC} > 40$ kpc in Fig. 11b might then perhaps not seem so
surprising, once we recall van den Bergh's (1995) remark that cluster
properties tend to correlate more strongly with their perigalactic distances
than with their current distances from the Galactic center. This shows the 
importance of obtaining accurate orbits for the Galactic GCs.

\begin{figure}[t]
\figurenum{11}
\plotfiddle{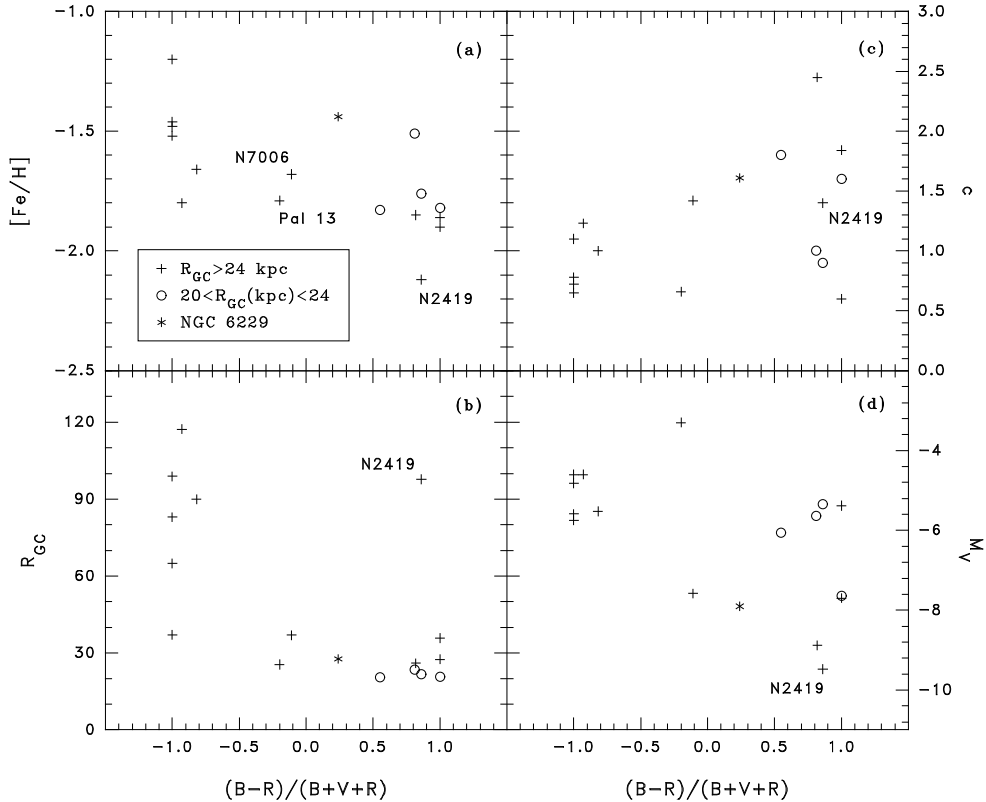}{3.825in}
{270}{80}{80}{-319.6}{400.0}
\caption{The HB morphology of outer-halo GCs, as a function of the metallicity
 (a); the Galactocentric distance (b), given in kpc; the central
 concentration $c$ (c); and the integrated visual magnitude $M_V$ (d).
 For NGC 5634, with $DT = 1$ (Harris 1996), we have assumed
 $(B-R)/(B+V+R) = +1$. Note that NGC 2419 is much more massive than
 the GCs with redder HB types lying at comparable galactocentric
 distances. }
\end{figure}

At any rate, based on an inspection of Table 7 and Fig. 11a, and ignoring for 
the moment the difference in masses among the outer-halo globulars, we would 
like to advance the suggestion that a prominent characteristic of the outer-halo
system of GCs, as far as the HB morphology is concerned, is the paucity
of objects with intermediate HB morphology types. In other words, we find a
hint that the HB morphology distribution of the {\em ensemble} of outer-halo GCs
seems to be bimodal itself. Apparently, the formation
of clusters with {\em either very red or very blue} HB types was favored
in the Galactic outer halo. NGC 7006 seems to be the only well-established
example of a distant ($R_{\rm GC} \geq 20$ kpc) cluster with a
more-or-less evenly populated HB ($B: V: R = 0.28: 0.33: 0.39$)
known to date. Pal 13 also has an intermediate
HB type, but the HB morphology is based on only 5 detected HB stars (Ortolani
et al. 1985; Borissova et al. 1996). NGC 6229, on the other hand,
has a very peculiar CMD, in the sense that it is the only outer-halo
cluster with a bimodal HB and an extended blue HB tail.
In fact, inspection of the published CMDs
for the outer-halo, blue-HB clusters in Table 7 (NGC 5694: Ortolani \& Gratton
1990; Pal 15: Harris 1991; NGC 2419: Christian \& Heasley 1988; NGC 5824:
Canon et al. 1990) confirms that none possesses as extended a blue HB
as does NGC 6229.

Figs. 11c and 11d also suggest a significant correlation between the HB
morphology and the central concentration $c$ and integrated luminosity
$M_V$. The correlations are definitely more significant than found by Lee et
al. (1994) in their Fig. 8. The possibility that the cluster environment may
affect the HB morphology has been discussed at considerable length by Fusi
Pecci et al. (1993, 1996), Djorgovski et al. (1991), Djorgovski \& Piotto
(1992), van den Bergh \& Morris (1993), and Buonanno et al. (1996).

Age estimates have also been published quite recently for several of the 
outer-halo globulars, allowing a test of the Searle \& Zinn (1978) hypothesis 
that the red HBs of some of these objects (cf. Table 7) are due to 
lower-than-average ages. In the studies of the CMD properties of Pal 4 
and Pal 14 by Christian \& Heasley (1986) and Holland \& Harris (1992), 
respectively, evidence was presented that they probably have ``normal" ages, 
in comparison with the clusters of the inner halo (but see Armandroff et al. 
1992 for a different point of view). This idea has received support in the 
recent investigation by Richer et al. (1996), who found that the ages of the 
outer-halo clusters Pal 3, Pal 4, NGC 2419, and NGC 6229 are very similar to 
those of the inner-halo objects in their sample. On the other hand, Sarajedini 
(1996) has recently challenged Holland \& Harris' results for Pal 14, and
Sarajedini \& Geisler (1996) have suggested that Pyxis is probably another 
example of a ``young" GC with a red HB. Thus, the Searle \& Zinn concept 
that the ages of the outer-halo clusters should be substantially lower than 
the ages of the inner-halo clusters remains very controversial.

NGC 6229 may potentially play an important r\^ole in deciding what parameter(s),
besides [Fe/H], control(s) the HB morphology for the outer-halo GCs. Since this
appears to be the only outer-halo cluster that contains {\em both
a blue and a red HB}, any successful explanation for the origin of this HB
bimodality might in principle (cf. Rood et al. 1993; Stetson et al. 1996)
also prove successful in explaining why the distribution of HB types for the
outer-halo globulars appears to be bimodal as well. As already noted, however,
the blue HB of NGC 6229 appears to be much more extended than in any other
outer-halo globular studied to date, which may perhaps suggest a physically
different origin for it, or an internal mechanism responsible for ``spreading"
the blue HB of NGC 6229 towards hotter temperatures (Buonanno et al. 1996).

Finally, we would like to add a caveat. From inspection of the orbital
properties of the outer-halo GCs, as recently provided by van den Bergh
(1993, 1995), we find that most of the clusters listed in Table 7 have
plunging orbits with respect to the Galactic center. Information on the
shape of the orbit is presently lacking for Pyxis and Pal 15. AM-1 and
NGC 6229 have uncertain orbital shapes. Also, the perigalactic distances
of most of these globulars appear to exceed 9 kpc (again, information on this
property is missing for the newly-discovered Pyxis cluster). NGC 6229, on
the other hand, seems to have a perigalactic distance as small as 4.2 kpc
(a similar value is found for NGC 5694 as well).
It should thus be exceedingly important to refine the orbital characteristics
for NGC 6229, so that its identification as a genuine member of the Galactic
outer halo can be either firmly established or ruled out. In either case,
NGC 6229, together with such clusters as NGC 1851 and NGC 2808, remains an
exciting challenge to stellar structure and evolution models.

\acknowledgments
J. B. and N. S. would like to thank  H. Markov for their help in the process
of obtaining the CCD frames, and also Drs. L. Georgiev, R. Kurtev, E. Chelebiev,
and T. Valtchev for their help. The authors gratefully acknowledge useful
comments and suggestions raised by Dr. S. Ortolani and Dr. Don VandenBerg
on earlier drafts of the manuscript, as well as the comments by an anonymous
referee. M. C. acknowledges useful discussions with S. Desidera and 
Dr. A. Sarajedini. This research was supported in part by the Bulgarian 
National Science Foundation grant under contract No. F-604/1996 with
the Bulgarian Ministry of Education and Sciences. This work
was performed while M. C. held a National Research Council--NASA/GSFC
Research Associateship. A. V. S. acknowledges support through NASA grant
NAG5-3028.

\clearpage
\singlespace

\end{document}